%% file: main.tex
\documentclass{article}
\usepackage{spconf,amsmath,graphicx}
\usepackage{adjustbox,multirow}
\usepackage{amssymb}
\usepackage{subfiles}
\usepackage{enumitem}
\usepackage[table,xcdraw]{xcolor}
\usepackage{url}
\usepackage{tabularx}
\usepackage{graphicx}
\usepackage[format=plain,textfont=it]{caption}
\usepackage{siunitx}
\usepackage{appendix}
\usepackage{balance}
\newcommand{\xmark}{\ding{55}}%

\let\origsection\section
\renewcommand\section[1]{\vspace{-0.25cm}\origsection{#1}\vspace{-0.1cm}}
\let\origsubsection\subsection
\renewcommand\subsection[1]{\vspace{-0.25cm}\origsubsection{#1}\vspace{-0.1cm}}
\let\origsubsubsection\subsubsection
\renewcommand\subsubsection[1]{\vspace{-0.25cm}\origsubsubsection{#1}\vspace{-0.1cm}}

\usepackage{amsmath,graphicx,amssymb}
\usepackage{xspace}
\usepackage{microtype, subfigure, verbatim}
\usepackage{hyperref}
\usepackage{booktabs, multicol, multirow, array}
\usepackage{arydshln, xcolor}
\usepackage{boldline,pifont}

\usepackage{array}
\newcolumntype{H}{>{\setbox0=\hbox\bgroup}c<{\egroup}@{}}
\title{\vspace{-0.45cm}Conformer-Based Self-Supervised Learning for Non-Speech Audio Tasks}

\name{
\begin{tabular}{c}
Sangeeta~Srivastava$^{\star \S}$, Yun~Wang$^{\dagger}$,
Andros~Tjandra$^{\dagger}$, 
Anurag~Kumar$^{\ddagger}$,
Chunxi~Liu$^{\dagger}$\\
Kritika~Singh$^{\dagger}$,
Yatharth~Saraf$^{\dagger}$
\end{tabular}
\thanks{$\S$ This work was done during an internship at Facebook.}
}
      
\address{$^\star$The Ohio State University, USA \quad
$^\dagger$Meta AI, USA \\ 
$^\ddagger$Reality Labs Research, USA \\ 
\texttt{\small srivastava.206@osu.edu, \{yunwang,androstj,anuragkr,chunxiliu,skritika,ysaraf\}@fb.com}}

\begin{document}
\ninept
\maketitle
\subfile{abstract}
\subfile{introduction}
\subfile{pretraining}
\subfile{downstream}
\subfile{table1}
\begin{figure}[t]
    \centering
	\includegraphics[width=0.95\columnwidth]{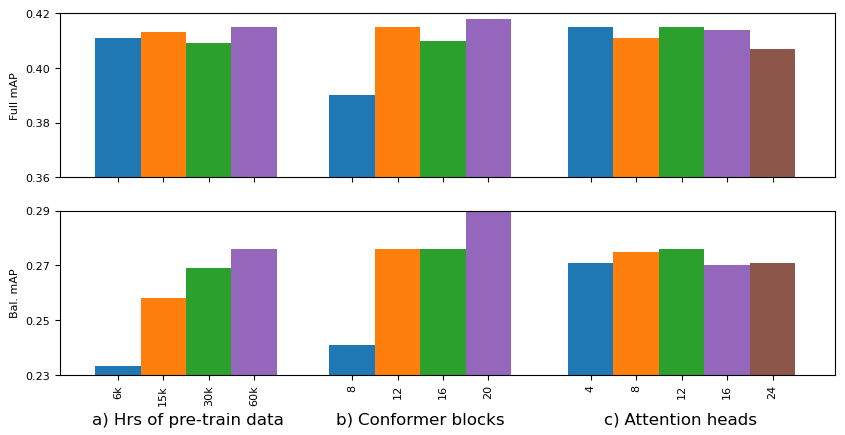}
    \vspace{-0.2cm}
	\caption{Effect of the amount of pre-training data, and the number of conformer blocks and attention heads (Full \& Balanced AudioSet).}
	\label{fig:ablation}
    \vspace{-0.2cm}
\end{figure}
\subfile{finetuning}
\subfile{table2}

\subfile{table3}
\subfile{results}
\subfile{conclusion}

%\balance
\bibliographystyle{IEEEbib-abbrev}
\bibliography{main}
\end{document}

%% file: abstract.tex
\begin{abstract}

\noindent Representation learning from unlabeled data has been of major interest in artificial intelligence research. While self-supervised speech representation learning has been popular in the speech research community, very few works have comprehensively analyzed audio representation learning for non-speech audio tasks. In this paper, we propose a self-supervised audio representation learning method and apply it to a variety of downstream non-speech audio tasks. We combine the well-known wav2vec 2.0 framework, which has shown success in self-supervised learning for speech tasks, with parameter-efficient conformer architectures. Our self-supervised pre-training can reduce the need for labeled data by two-thirds. On the AudioSet benchmark, we achieve a mean average precision (mAP) score of 0.415, which is a new state-of-the-art on this dataset through audio-only self-supervised learning.
Our fine-tuned conformers also surpass or match the performance of previous systems pre-trained in a supervised way on several downstream tasks. We further discuss the important design considerations for both pre-training and fine-tuning.

%The most common way to learn representations for non-speech audio in a self-supervised manner is pre-training with multimodal data. Nevertheless, pre-training with audio-only data can be more privacy friendly. We propose to use the wav2vec 2.0 framework for self-supervised pre-training on non-speech audio data, considering that this framework has been successful in learning useful representations for speech. We employ the conformer architecture instead of the original transformer, because conformers are more parameter-efficient. On the AudioSet benchmark, we achieve a new state-of-the-art mean average precision score of 0.415 in the category of self-supervised learning using only audio. By evaluating our self-supervised learning scheme on a diverse set of classification tasks, including but not limited to acoustic event detection, we show that it can close the gap between self-supervised and supervised pre-training on multiple datasets. We further discuss the important design considerations for both pre-training and fine-tuning.
\end{abstract}

\vspace{-0.13cm}

\begin{keywords}
Self-supervised learning, representation learning, conformer, sound events, wav2vec.
\end{keywords}

%% file: introduction.tex
\section{Introduction}
\label{sec:intro}

In recent years, self-supervised learning (SSL) has proven to be an effective method for representation learning without the need for labeled supervision, not only in the language \cite{devlin2018bert} and visual \cite{chen2020simple} domains but also in the audio domain. In the audio domain, SSL for speech tasks have shown extremely promising results \cite{baevski2020wav2vec}, but SSL for non-speech audio tasks has been explored to a lesser extent. Examples of SSL for non-speech tasks include multimodal self-supervised learning \cite{akbari2021vatt} and audio captioning \cite{liu2021cl4ac}, but an extensive evaluation of SSL for non-speech audio-only tasks remains to be seen.

A number of self-supervised representation learning frameworks have been proposed, of which contrastive learning based methods have been found to be particularly effective in both image and speech domains. It aims to push semantically comparable samples closer together and dissimilar samples apart in the feature space. Wav2vec 2.0 \cite{baevski2020wav2vec} is one such contrastive learning framework which have been demonstrated to be highly efficient in speech-related tasks \cite{tjandra2021improved, Fan2021ExploringW2}.

Wav2vec 2.0 uses transformers to build contextualized representations of the speech sequence, but transformers can only simulate global dependencies. While convolutional networks (CNN) can capture local associations, obtaining global information from the model would necessitate the use of additional layers or parameters. Conformers \cite{gulati2020conformer}, on the other hand, captures both local and global dependencies while requiring fewer parameters as compared to convolutional models. Prior works \cite{zhang2020pushing, miyazaki2020convolution} have shown that conformers can outperform transformers and CNN models on both speech and non-speech tasks. 

In this work, we combine the conformer architecture with contrastive learning to learn representations for non-speech audio in a self-supervised manner. We replace the transformer in wav2vec 2.0 with a conformer in order to capture both global and local information in the audio signal. Unlike prior self-supervised works on speech representation learning (\textit{e.g.}~\cite{baevski2020wav2vec}) which directly learn from raw audio waveforms, we use logmel spectrograms as the fundamental representation of the audio signal. Besides their proven efficacy in different audio tasks \cite{kumar2021sound}, logmels are much more compact and lead to better memory and compute efficiency. The network is pre-trained on a large-scale unlabeled dataset of $\sim$67,000 hours to learn audio representations, and then fine-tuned on several audio tasks to show the generalization capabilities of the proposed SSL framework. 

The contributions of this work are the following:
\textbf{(1)} We propose a conformer-based SSL framework for general-purpose audio representation learning applicable to many audio tasks, which can reduce the need for labeled data by two-thirds;
\textbf{(2)} On the AudioSet \cite{gemmeke2017audio} benchmark, our fine-tuned representations achieve a new state-of-the-art (SOTA) mean average precision (mAP) of 0.415 for self-supervised learning with only audio, outperforming the previous best score of 0.329;
\textbf{(3)} Our method can surpass or match the performance of previous systems pre-trained in a supervised way on 3 out of 4 classification tasks: acoustic scenes, music and human actions;
\textbf{(4)} We show the effect of the design choices during pre-training and identify the main parameters to tune to avoid overfitting during fine-tuning.

%% file: pretraining.tex
\section{Self-Supervised Conformer Training}
\begin{figure*}
    \centering
	\includegraphics[width=0.85\textwidth, trim = 0cm 3cm 0cm 4.5cm, clip]{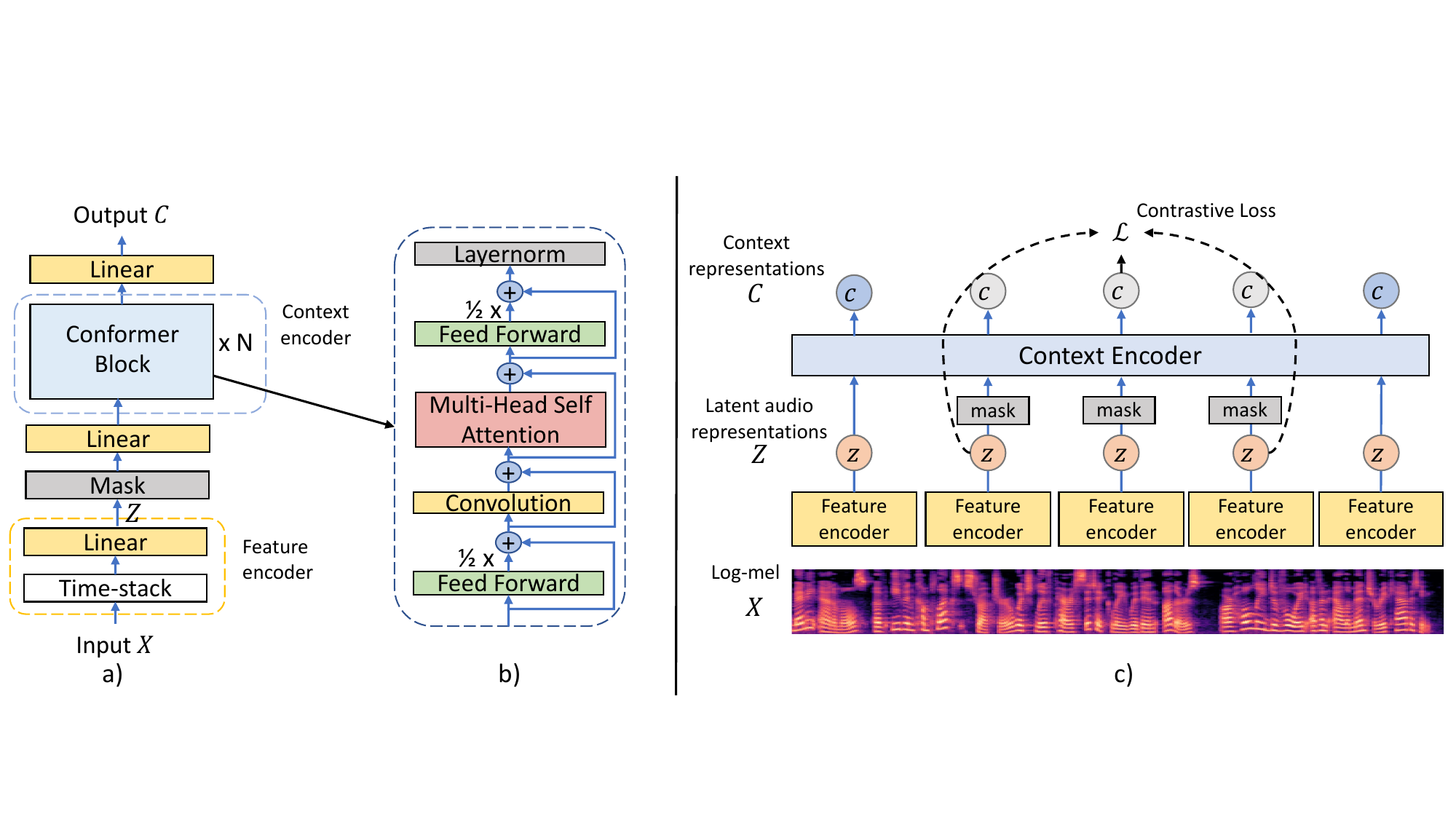}
	\caption{a) Logmel wav2vec 2.0 conformer architecture. b) A conformer block. c) An illustration of how wav2vec 2.0 generates context representation and solves the contrastive task.}
	\label{fig:wav2vec2}
	\vspace{-0.3cm}
\end{figure*}

\vspace{0.25cm}

\subsection{Upstream Data}
%We use an in-house large scale unlabeled audio dataset for self-supervised pre-training of the conformer networks. The dataset consists of audio recordings from $\sim$3.9M videos totaling $\sim$67k hours. The audio recordings are up to 5 minutes long and were filtered from a much larger dataset which an audio event detection system~\cite{wang2019comparison}. The filtering is done to ensure the unlabeled dataset contains exhaustive acoustic signatures which will ensure that the network learns from an generalizable dataset. 

The data for pre-training was selected from de-identified audio tracks of publicly shared Facebook user videos. We ran an in-house version of TALNet~\cite{wang2019comparison} on the audio tracks of 440 million user videos up to five minutes long, and obtained estimated probabilities of the 527 types of acoustic events in the AudioSet ontology. Then, for each event type, we selected 9,500 videos with the highest probabilities (some videos might be selected for multiple event types). The resultant dataset contains $\sim$3.9M videos totaling $\sim$67k hours. We expect that this large dataset captures enough relevant acoustic phenomena and is sufficiently diverse for different non-speech audio tasks.

\subsection{Architecture}
\label{sec:architecture}

\noindent \textbf{Frontend.} 
We use 64-dimensional logmel spectrograms extracted with a window size of 64 ms and a hop size of 20 ms as the input $X$.
%to the wav2vec 2.0 framework. 
%This differs from the original wav2vec setup, which uses audio waveforms as the input. 

\noindent \textbf{Encoder.} As shown in Figure \ref{fig:wav2vec2},
the logmel spectrograms are fed into two types of encoders: feature and context encoders. In the feature encoder, the time stacking layer stacks every $R=4$ consecutive frames into a single frame; then a linear layer converts the stacked spectrogram into latent representation vectors $Z=[z_1,...,z_{T/R}]$.

Before feeding the latent representation $Z$ to the context encoder, we mask a proportion of it, similar to the original wav2vec 2.0 framework. The context encoder consists of a linear layer, multiple conformer blocks \cite{gulati2020conformer} and another linear layer at the end. 
% \COMMENT{@Chunxi: Can you add some implementational details of conformers if they are any different from the original paper?}  -- yes, i have edited the following paragraph. Also please "swap the order of the convolution module and the multi-head self-attention module " in figure 1 a).
We experiment with two context encoder architectures of differing sizes: 
%  which mainly differ in the conformer hyperparameters:  \textbf{conformer small} (cf\_S) and \textbf{conformer large} (cf\_L). 
\vspace{-0.1cm}
\setlength{\itemsep}{0pt}
\begin{enumerate} [label=(\roman*)]   
\item \textbf{conformer small} (\textit{cf\_S}): 12 conformer blocks with 256-D encoder embeddings and 8 attention heads (18.4M parameters).
\item \textbf{conformer large} (\textit{cf\_L}): 12 conformer blocks with 768-D encoder embeddings and 12 attention heads (88.1M parameters).
\end{enumerate}
\vspace{-0.1cm}
Both encoders use a feed-forward network (FFN) dimension of 1024, dropout 0.1, kernel size 31 in the first conformer block and 15 for the rest.
We further explore different design choices in Section~\ref{sec:ablation}.
Additionally, following \cite{li2021better}, we remove the original relative positional encoding, and reuse the existing convolution module for positional encoding by swapping the order of the convolution and multi-head self-attention modules, which speeds up both training and inference.  

\noindent \textbf{Contrastive Loss.}
We learn audio representations by solving a contrastive task: identify the true latent representation $z_t$ for a masked time step $t$ within a set of $K + 1$ candidates $\tilde{Z} = \{\tilde{z}_1, \ldots, \tilde{z}_{K+1}\}$, which includes $z_t$ and $K$ distractors. Similar to the original wav2vec 2.0, the distractors are uniformly sampled from other masked time steps of the same audio snippet. Given the context network output $c_t$ centered over a masked time step $t$, the contrastive loss is defined as:
\begin{equation}
L = -\log{\frac{\exp(\mathrm{sim}(c_t,z_t))}{\sum_{\tilde{z} \in \tilde{Z}}\exp(\mathrm{sim}(c_t,\tilde{z}))}}
\end{equation}
%\begin{equation}
%    \mathrm{sim}(\mathbf{x}, \mathbf{y}) = \frac{\mathbf{x}^T \mathbf{y}}{\norm{\mathbf{x}} \norm{\mathbf{y}}}
%\end{equation}
where $\mathrm{sim}(\mathbf{x}, \mathbf{y})$ is the cosine similarity.

\subsection{Hyper-parameters}
For all pre-training experiments, we use the Adam \cite{kingma2014adam} optimizer with $\beta_1 = 0.9, \beta_2 = 0.98$. We warm up the learning rate for the first 10k updates linearly to a peak of $3 \times 10^{-4}$, and then decay it linearly to zero until 300k updates. We regularize the model using weight decay with a decay factor of 0.01. For the contrastive loss we use $K = 100$ distractors and mask 30\% of the latent representations $Z$. 

%% file: downstream.tex
\section{Downstream Tasks}

We evaluate the generalizability of the learnt representations on sound event detection (SED) and other non-speech audio  tasks. Within SED, we evaluate on the AudioSet dataset separately in Sec.~\ref{sec:results}, because it is much larger than the data used in all the other downstream tasks, and has been used in several priors works as well. % Moreover, while we only fine-tune on AudioSet, several prior works \cite{akbari2021vatt} pre-train as well as fine-tune on AudioSet.  

\subsection{Sound Event Classification}

\textbf{AudioSet}~\cite{gemmeke2017audio}: AudioSet contains $\sim$ 2 million 10-second audio clips from YouTube videos, labeled using an ontology of 527 sound classes. Each clip may have multiple labels. The \emph{Balanced} training set of AudioSet is a subset with at least 59 clips per class. The balanced training, full training, and evaluation sets contain 22k, 2M, and 20k samples, respectively. For AudioSet, we use the larger model (\textit{cf\_L}).

\noindent \textbf{ESC-50}~\cite{piczak2015esc}: ESC-50 consists of 2,000 5-seconds audio clips belonging to 50 environmental sound categories. It contains 40 samples for each category, and comes divided into 5 folds for cross validation.

\noindent \textbf{FSDKaggle2019}~\cite{fonseca2019audio}: FSDKaggle2019 is a multi-label dataset consisting of 29,266 audio files annotated with 80 labels form the AudioSet ontology. It further consists of a \textit{curated} set and a \textit{noisy} set: the former is annotated by humans but the labels may be incomplete; the latter contains many wrong labels.

\subsection{Other Audio Tasks: Scenes, Music, and Human Actions}

\textbf{Acoustic scene classification}: We use the dataset from Task 1a of the 2019 DCASE challenge \cite{mesaros2018multi}. It consists of 9,185 and 4,185 segments in the training and the test sets respectively, belonging to 10 acoustic scene classes such as ``airport'', ``park'', and ``metro station''.
    % The data consists of recordings from ten cities; the training subset contains recordings from only 9 of the cities, to test the generalization properties of the model. 
    
\noindent \textbf{Music tagging}: This is a multi-label classification problem where each recording can have one genre label and multiple instrument labels. For this task, we use the MagnaTagATune \cite{law2009evaluation} dataset which consists of 25,863 music clips, each clip lasting 29 seconds. We follow the most common 12:1:3 split of training, validation, and evaluation data, and only use the 50 most popular tags out of the 188.
    
\noindent \textbf{Human action classification}: This task involves recognizing human actions such as ``dancing'' and ``playing guitar'' in video recordings. The Kinetics700 dataset \cite{kay2017kinetics} is a collection of 650k 10-second video clips covering 700 human action classes. This problem has primarily been addressed using images in a uni- \cite{feichtenhofer2021large} or multimodal \cite{radford2021learning} approach; so far there is only one audio-only solution~\cite{kumar2021sound}.

%% file: table1.tex
\begin{table}[t]
\centering
\setlength{\tabcolsep}{0.35em}
\resizebox{\columnwidth}{!}{
\begin{tabular}{c|c|c|c|c|c}
\hlineB{2}
& \textbf{Balanced} & &  & & \textbf{Full} \\
\textbf{Model} & \textbf{(20k)} & \textbf{60k} & \textbf{200k} & \textbf{600k} & \textbf{($\sim$2M)} \\
\cline{2-6}
 & 1\% & 3\% & 10\% & 30\% & 100\% \\
\hline
From scratch & 0.138 & - & - & - & 0.366 \\ \hline
Pre-train + fine-tune & \textbf{0.276} & 0.305 & 0.337 & 0.356 & \textbf{0.415} \\
\hlineB{2}
\end{tabular}
}
\caption{Performance (measured in mAP) of the \textit{cf\_L} model on the test set of AudioSet, with and without pre-training, and with different amounts of fine-tuning data.}
\label{table:audioset}
\vspace{-0.2cm}
\end{table}

%% file: finetuning.tex
\section{Fine-tuning}

After the self-supervised pre-training, the conformer model can generate an embedding vector (the context representation) for each frame of the input audio. The downstream tasks, however, require predictions over the entire input audio. For AudioSet, we first add a linear classification layer with the sigmoid activation to predict the frame-level probabilities of each event type, then add a linear softmax pooling layer~\cite{wang2019comparison} to pool them into global probabilities. For all other downstream tasks, we first average-pool the embeddings across all the frames, then stack a linear classification layer to make predictions.

During fine-tuning, we allow the weights of the conformer to update. We find this is critical to make the SSL schema work. This is different from some existing works on transfer learning, where a shallow classifier is stacked on top of an encoder with frozen weights. 

\subsection{Data Processing}

%The AudioSet dataset, in particular, has a highly skewed label distribution. To address this problem, we apply a data balancing algorithm when forming minibatches. A data sampler for each event and a meta-sampler that cycles through the samplers ensure that the training sees different sound classes at approximately equal frequencies. 

%and we apply To ensure the model sees instances of different event types at approximately equal frequencies, we apply the following data balancing algorithm when sampling data to form minibatches. We create one sampler for each event type, which cycles through the videos containing that event and shuffles the videos after each pass. We also create a meta-sampler that cycles through the samplers, and shuffles the samplers after each pass. Each query on the meta-sampler will yield a single instance of a random event type.

To deal with the highly skewed label distribution in Full AudioSet, we apply data balancing to ensure that the training sees different sound classes at approximately equal frequencies.
We also apply data augmentations in the following order to all downstream tasks:

\noindent \textbf{Temporal jittering}: Prior to extracting logmel features, we shift the waveform by a random amount between -200 and 200 samples. The maximum shift (200 samples) is equal to half the frame length.

\noindent \textbf{SpecAugment}~\cite{park2019specaugment}: For each 10~s video, we mask out the logmel features of one random temporal interval up to 2~s. We do not mask out any frequency bins because it does not work well with Mixup.

\noindent \textbf{Mixup}~\cite{zhang2017mixup}: For a minibatch of $n$ videos, with logmel features denoted by $\{\mathbf{x}_1, \ldots, \mathbf{x}_n\}$, we shuffle the instances to get $\{\mathbf{x}'_1, \ldots, \mathbf{x}'_n\}$, and then mix up the two batches according to
\begin{equation}
\mathbf{x}''_i = \alpha_i \mathbf{x}_i + (1 - \alpha_i) \mathbf{x}'_i, \quad i = 1, \ldots, n
\end{equation}
The mixing coefficient $\alpha_i$ is sampled from the beta distribution $B(0.5, 0.5)$, and it is enforced that $\alpha_i \ge 1 - \alpha_i$. The labels of $\mathbf{x}''_i$ inherits those of $\mathbf{x}_i$; we do not mix them with the labels of $\mathbf{x}'_i$.

\subsection{Consistency Loss}
In addition to the binary cross-entropy (BCE) loss, we employ a consistency loss as a regularizer for training. For each minibatch, we apply the data augmentation twice with different random numbers, and compute the symmetric KL divergence between the model's predicted event probabilities on the two versions of data. This KL divergence is multiplied by 2 and added to the BCE loss.

\subsection{Hyperparameter Tuning}
We find that the fine-tuning process is prone to overfitting, but it can be avoided by carefully tuning the learning rate, the learning rate schedule, and the batch size.
%Learning rate, learning rate schedule and batch size play important role in the learning process and careful tuning of these hyperparameters can help avoid overfitting.
We utilize a three-stage learning rate scheduler \cite{park2019specaugment} for all downstream datasets. The three stages are: warmup, hold, and exponential decay, and they typically last for 30\%, 30\%, and 40\% of all the updates. While high-resource datasets perform best with larger batches, smaller datasets prefer smaller batches. For example, we use a batch size of 640 and 64 for Full AudioSet and MagnaTagATune, respectively. We also regularize the model more for low-resource datasets by adding dropout to the output layer of the pre-trained conformer. We sweep different combinations of the parameters to find the one that optimizes validation performance.

%% file: table2.tex
\begin{table}[t]
\centering
\setlength{\tabcolsep}{0.35em}
\resizebox{\columnwidth}{!}{
\begin{tabular}{c|c|c|c|c|c}
\hlineB{2}
& & & \textbf{AudioSet} & \textbf{Fine-} & \textbf{Test} \\
\textbf{Pre-training} & \textbf{Model} & \textbf{Input} & \textbf{pre-train?} & \textbf{tune?} & \textbf{mAP} \\ 
\hline \hline
& C$^3$ \cite{jansen2020coincidence} & l & \checkmark & & 0.206 \\
& Triplet \cite{jansen2018unsupervised} & l & \checkmark & & 0.259 \\
Audio-only & CPC \cite{wang2020contrastive} & w & \checkmark & & 0.277  \\
Self-supervised & Multi-Format \cite{wang2021multi} & w + l & \checkmark & & 0.329 \\
& Separation-based \cite{fonseca2021self} & l & \checkmark & & 0.326 \\
% \rowcolor[HTML]{FFCC67} 
& \textbf{Ours} (\textit{cf\_L}) & \textbf{l} & \xmark & \textbf{\checkmark} & \textbf{0.411} \\ 
\hline
& C$^3$ \cite{jansen2020coincidence} & l + v & \checkmark & & 0.285 \\
Multimodal & MMV \cite{alayrac2020self} & l + v + t & \checkmark & & 0.309 \\
Self-supervised & VATT \cite{akbari2021vatt} & l + v + t & \checkmark & \checkmark & 0.394  \\
& Multimodal COLA \cite{wang2021multimodal} & w + l + v & \checkmark & & 0.424 \\ 
\hline
Supervised & PSLA (w/o ensemble) \cite{gong2021psla} & w + l & & \checkmark & 0.444 \\
(ImageNet) & AST (w/o ensemble) \cite{gong2021ast}  & l & & \checkmark & 0.459 \\
\hline
& TALNet \cite{wang2019comparison} & l & &  & 0.383 \\
No pre-training & WEANet-SUSTAIN \cite{kumar2020sequential} & l & &  & 0.398 \\
(From scratch) & PANNs \cite{kong2020panns} & w + l & &  & 0.431 \\
& ERANN \cite{verbitskiy2021eranns} & l & &  & 0.450 \\
\hlineB{2}
\end{tabular}
}
\caption{\textbf{Performance comparison with prior works on Full AudioSet.} Input forms include waveforms (w), logmel (l), video (v), and text (t). We label systems that are pre-trained on AudioSet, or update the encoder weights during fine-tuning. Higher mAP is better.}
\label{table:audioset-others}
\vspace{-0.3cm}
\end{table}

%% file: table3.tex
\begin{table*}[t]
\centering
\resizebox{\textwidth}{!}{
\begin{tabular}{c|c|c|c|c|c|c|c|c|c}
\hlineB{2}
\multirow{2}{*}{\textbf{Task}} & \multicolumn{2}{c|}{\multirow{2}{*}{\textbf{Dataset}}} & \textbf{\# Training} & \multirow{2}{*}{\textbf{Metric}} & \multicolumn{2}{c|}{\textbf{SS Conformer}} & \multicolumn{3}{c}{\textbf{Prior Works}} \\ \cline{6-10} 
 &  \multicolumn{2}{c|}{} & \textbf{Samples} & & \hspace{0.03cm} \textbf{cf\_S} \hspace{0.03cm} & \textbf{cf\_L} & \textbf{SS + shallow} & \textbf{S + shallow} & \textbf{S + fine-tune}
 \\ \hline \hline
\multirow{3}{*}{\textbf{Sound Events}} & \multicolumn{2}{c|}{ESC50} & \phantom{00}1,600 &  Accuracy & 80.7 & 88.0 & 86.3 \cite{wang2021multi}$^\dagger$ & 94.1 \cite{kumar2021sound} & 94.7 \cite{kong2020panns} \\ \cline{2-10} 

 & \multirow{2}{*}{FSDKaggle} & curated & \phantom{00}4,970 & \multirow{2}{*}{lwlrap} & 50.3 & 61.1 & \xmark & 72.8 \cite{kumar2021sound} & \xmark \\ \cline{3-4} \cline{6-10} 
 &  & noisy & \phantom{0}19,185 &  & 33.2 & 40.1 & \xmark & 51.0 \cite{kumar2021sound} & \xmark \\ \hline
\textbf{Acoustic Scenes} & \multicolumn{2}{c|}{DCASE2019 Task 1a} & \phantom{00}9,185 & Accuracy & 72.4 & 76.1 & \xmark & 68.0 \cite{kumar2021sound} & \xmark \\ \hline

\textbf{Music Tagging} & \multicolumn{2}{c|}{MagnaTagATune} & \phantom{0}15,247 & MAUC & 90.2 & 91.2 & \multicolumn{1}{c|}{89.3 \cite{spijkervet2021contrastive}} & 91.5 \cite{kumar2021sound} & \xmark \\ \hline
\textbf{Human Actions} & \multicolumn{2}{c|}{Kinetics700} & 504,443 & Top-1 Accuracy & 20.4 & 23.5 & \xmark & 18.0  \cite{kumar2021sound} & \xmark \\
\hlineB{2}
\end{tabular}
}
\caption{Performance of both versions of our conformer on downstream tasks, compared with prior works using either self-supervised (SS) or supervised (S) pre-training. Most prior works use shallow classifiers; only \cite{kong2020panns} uses fine-tuning as we do. All metrics are the higher, the better.
$^\dagger$Performance using logmel representations only. Combining waveform and logmel features produces a higher accuracy of 90.5\%.}
\label{table:performance}
\vspace{-0.2cm}
\end{table*}

%% file: results.tex
\section{Results}
\label{sec:results}

\vspace{0.25cm}

\subsection{Evaluation on AudioSet}

\vspace{0.25cm}

\subsubsection{Effect of Pre-training}

First, we evaluate the merit of pre-training in our proposed SSL schema in Table~\ref{table:audioset}. On both the Balanced and the Full training sets, we compare the performance of the \textit{cf\_L} model trained from scratch vs pre-trained then fine-tuned. The model trained from scratch performs \textbf{50\%} and \textbf{12\%} worse than the pre-trained and fine-tuned version on Balanced and Full AudioSet, respectively. This emphasizes the necessity of pre-training, especially for smaller downstream datasets.

The merit of pre-training can also be quantified by the reduced need for labeled data. By varying the amount of fine-tuning data from 20k to 1.9M audio clips, we find that pre-training and fine-tuning with 600k clips approximately matches the performance of training from scratch with 1.9M clips. That is, pre-training with $\sim$67k hours of unlabeled data reduces the need for labeled data by \textbf{two-thirds}.

\subsubsection{Ablation Studies}
\label{sec:ablation}

We perform detailed ablation studies to investigate the effect of the amount of pre-training data and the design choices regarding the model structure, on both the Full and and Balanced training sets.

\textbf{Amount of pre-training data}. As shown in Fig.~\ref{fig:ablation}a, the performance on Full AudioSet stays relatively unchanged as we vary the amount of pre-training data from 6k to 60k hours. In other words, we can reduce the need for labeled data by two thirds even with only 6k hours of unlabeled data. On Balanced AudioSet, however, the performance increases significantly and monotonically with every bit of extra pre-training data. This demonstrates that a sufficient amount of unlabeled data can make up for the scarcity of labeled data.

\textbf{Number of conformer blocks}. We vary the depth of the model from 8 conformer blocks (58.9M parameters) to 20 conformer blocks (146.6M parameters), keeping all other hyperparameters identical to the large model. Fig.~\ref{fig:ablation}b shows that the largest performance gain occurs when going from 8 to 12 blocks. Beyond that, the improvement is minimal (if any) for Full AudioSet, but still moderate for Balanced AudioSet from 16 to 20 blocks.

\textbf{Number of attention heads}. Attention heads allow varied levels of focus on different parts of the sequence, producing better predictions than a single weighted average. We vary the number of attention heads from 4 to 24 in our large model, using the same number of heads in all layers. As shown in Fig.~\ref{fig:ablation}c, there is a slight improvement from 4 to 12 heads, but the performance drops once the number of attention heads exceeds 12 for both Full and Balanced AudioSet.

In general, the pre-training data size and design choices have a larger impact when labeled data for the downstream task is limited.

\subsubsection{Comparison with Prior Works}

Table~\ref{table:audioset-others} lists the Full AudioSet test performance of our system and some prior works. Depending on the type of data used for pre-training, we divide the works into four categories: self-supervised pre-training with audio only, self-supervised pre-training with multimodal data, supervised pre-training, and no pre-training. Not all comparisons are apple to apple: some works pre-train on AudioSet, and others (including ours) fine-tune the weights of the entire network, both of which may give them an advantage. Still, we report our performance with $\sim$6k hours of pre-training data without any augmentation, in order to match previous works that pre-train on AudioSet.

%Table~\ref{table:audioset-others} compares our test performance on AudioSet with prior works. We divide the prior works into three groups: self-supervised (SS) with audio only, self-supervised with multimodal signals, and supervised (S) learning methods. Our work falls strictly in the first category. To be fair to the previous works, we compare our model pre-trained on $\sim$6k hours of in-house data without any augmentation to match the amount of pre-training data used by these previous works. Note that several of these works \cite{wang2020contrastive, wang2021multi, akbari2021vatt} pre-train on AudioSet itself, which gives them an advantage with respect to performance on AudioSet.

%Note that, despite keeping a similar amount of pre-training data, there might be differences in the experimental setup. For example, \cite{kumar2021sound} uses feature-based learning whereas we use fine-tuning; \cite{fonseca2021self} uses an embedding dimension of 1,024 whereas we use 768; \cite{wang2021multi} uses a window size of 20~ms and a stride of 10~ms, unlike our choice of 64~ms window size and 20~ms stride. 

We give a quick overview of the previous works using audio-only self-supervised pre-training. \cite{wang2020contrastive} uses contrastive predictive coding (CPC) to single out the representation of a future step at a given distance, but also adds a nonlinear learnable similarity metric and adversarial perturbation. All the others~\cite{jansen2020coincidence, jansen2018unsupervised, wang2021multi, fonseca2021self} learn encoders such that a pair of audio clips sampled within a certain temporal proximity have similar representations; the difference lies in how the pairs are presented to the encoder, and the loss function. \cite{jansen2020coincidence} and \cite{jansen2018unsupervised} present both clips as spectrograms, and use the BCE loss and triplet loss, respectively. \cite{wang2021multi} presents one clip as a waveform and the other as a spectrogram, and uses a one-vs-many contrastive loss. \cite{fonseca2021self} separates a mixed signal into two channels and forms a pair using one of the channels and the original signal, and combines the BCE and the one-vs-many contrastive losses.

The current state-of-the-art performance on AudioSet using audio-only SSL is held by \cite{wang2021multi}, with an mAP of 0.329. Our method outperforms it by a huge \textbf{25\%} relative, and achieves an mAP of 0.411. However, it should be pointed out that we update encoder weights during fine-tuning, while all the existing works in this category, including \cite{wang2021multi}, use a shallow classifier upon frozen encoder weights.

On the multimodal self-supervised learning front, most methods use audio and visual signals to learn relationships between them \cite{jansen2020coincidence, alayrac2020self, akbari2021vatt, wang2021multimodal}. In addition to audio and video, some works \cite{alayrac2020self, akbari2021vatt} also make use of textual information, but perform worse than our framework with only audio. Our audio-only learning method is competitive even to the best multimodal SSL work on AudioSet~\cite{wang2021multimodal}, inferior by only 3\% relative.

Supervised pre-training on ImageNet~\cite{gong2021psla,gong2021ast} outperforms our proposed SSL framework, but it requires a large amount of labeled data for pre-training. However, we notice some systems trained from scratch~\cite{kong2020panns,verbitskiy2021eranns} achieve exceedingly high performance. This may indicate that CRNNs may be better suited for SED than conformers.

\vspace{-0.2cm}
\subsection{Evaluation on Other Downstream Tasks}
Table \ref{table:performance} summarizes the results for all the other downstream tasks. We compare the performance of both our conformer models (\textit{cf\_S} and \textit{cf\_L}) with prior works of transfer learning using either self-supervised or supervised pre-training. Most of these prior works use shallow classifiers; only [34] fine-tunes the entire network as we do. To our knowledge, no self-supervised prior work exists for FSDKaggle2019, DCASE2019, or Kinetics700.

For ESC50 and MagnaTagATune, we outperform baselines of self-supervised pre-training with our \textit{cf\_L} architecture. For music tagging, we nearly match the baseline system~\cite{kumar2021sound} pre-trained in a supervised fashion; for the classification of acoustic scenes and human actions, \textit{cf\_L} (\textit{cf\_S}) outperforms supervised pre-training baselines by 11.9\% (6.5\%) and 30.5\% (13.3\%), respectively. However, if we compare against supervised pre-training baselines for SED tasks, our best \textit{cf\_L} architecture performs 7.6\% (ESC50) and 19.1\% (FSDKaggle curated) worse. This significant disparity can be attributed to the fact that the baseline systems were pre-trained on Full AudioSet, which has a substantial overlap in the label space with both ESC50 and FSDKaggle, providing the baseline systems with an edge.

%% file: conclusion.tex
\vspace{-0.3cm}
\section{conclusion}
\vspace{-0.1cm}

In this paper, we have proposed and evaluated a conformer-based self-supervised framework for learning representations for non-speech audio. The self-supervised pre-training can reduce the need for labeled data by two-thirds. On the widely known AudioSet, our framework produces an mAP of 0.415, outperforming the audio-only self-supervised learning SOTA of 0.329; it also achieves performance comparable to supervised pre-trained baselines on several other downstream non-speech tasks. Our ablation studies shows the significance of critical design choices such as the model depth. 

%With ablation studies, we have shown that the model depth and amount of pre-training data play a critical role when labeled data is limited. We have also pointed out the important hyperparameters to tune to avoid overfitting during fine-tuning.